\documentclass[a4paper,12pt]{article}
\usepackage[dvips]{graphicx}
\def\gev{{\rm GeV}}
\def\mev{{\rm MeV}}

\def\etal{{\it et al.}}

\newcommand{\beq}{\begin{equation}}
\newcommand{\eeq}{\end{equation}}
\newcommand{\bea}{\begin{eqnarray}}
\newcommand{\eea}{\end{eqnarray}}
\newcommand{\bsub}{\begin{subequations}}
\newcommand{\esub}{\end{subequations} \noindent}

\def\PRD#1#2#3{Phys. Rev. {\bf D#1} (#2) #3}

\def\NPB#1#2#3{Nucl. Phys. {\bf B#1} (#2) #3}
\def\PTP#1#2#3{Prog. Theor. Phys. {\bf #1} (#2) #3}

\def\PLB#1#2#3{Phys. Lett. {\bf B#1} (#2) #3}
\def\PRL#1#2#3{Phys. Rev. Lett. {\bf #1} (#2) #3}

\catcode`\@=11 
\def\lsim{\mathrel{\mathpalette\@versim<}}
\def\gsim{\mathrel{\mathpalette\@versim>}}
\def\@versim#1#2{\vcenter{\offinterlineskip
        \ialign{$\m@th#1\hfil##\hfil$\crcr#2\crcr\sim\crcr } }}
\catcode`\@=12 
\begin{document}
\begin{titlepage}
\begin{flushright}
\begin{tabular}{l}
{\bf OCHA-PP-197}\\
{\bf hep-ph/0212242}
\end{tabular}
\end{flushright}
\begin{center}
    \vspace*{1.2cm}
    
    {\Large\bf Neutrino Mass Matrix  \\
          Out of Up-Quark Masses Only}
    \vspace{1.5cm}
    
    {\large
      Masako {\sc Bando}\footnote{E-mail address:
        bando@aichi-u.ac.jp} 
      and Midori {\sc Obara}\footnote{E-mail address:
        midori@hep.phys.ocha.ac.jp}}   \\
    \vspace{7mm}
    $^*$ {\it Aichi University, Aichi 470-0296, Japan} \\[1mm]
    $^{\dagger}$ {\it Graduate School of Humanities and Sciences, \\ 
Ochanomizu University, Tokyo 112-8610, Japan}
\end{center}
    \vspace{1.5cm}
\begin{abstract}
\noindent
Under the assumption of 
symmetric four zero texture for fermion mass matrices in $SO(10)$ model, 
the neutrino Dirc mass matrix is derived from the up-quark mass 
matrix. By adjusting two scales of the Majorana mass sector, we 
can derive observed neutrino two large mixing angles 
very naturally. It is indeed remarkable that 
all the masses and mixings of neutrinos are 
expressed in terms of  
only three known parameters, $m_t,m_c,m_u$. 
A typical example of our model is 
\bea
\tan^2 2\theta_{\tau \mu}
=\frac{4}{(1-2\frac{m_c}{\sqrt{m_u m_t}})^2}. \nonumber
\eea
\end{abstract}
\end{titlepage}

Recent neutrino experiments by Super-Kamiokande 
\cite{kamioka:2002pe,kamioka:2001nj} and SNO~\cite{SNO} 
have confirmed  
neutrino oscillations with large mixing angles:
\begin{eqnarray}
     \sin^22\theta_{atm}   &>&0.83 \,(99 \%~C.L.),    \nonumber \\ 
     \tan^2 \theta_{sol}   &=& 0.24-0.89 \,(99.73 \%~C.L.),
\label{expangle}
\end{eqnarray}
with mass squared differences of
\begin{eqnarray}
     \Delta m^2_{atm}   &\sim& 2.5 \times 10^{-3}~\rm{eV}^2,    \\ 
     \Delta m^2_{sol}   &\sim& 5 \times 10^{-5}~\rm{eV}^2. 
\end{eqnarray}

The neutrino mixing angles are expressed by MNS matrix~\cite{MNS} which 
is written as
\begin{equation}
V_{MNS}=U_l^{\dagger}U_{\nu},
\label{MNS}
\end{equation}
with $U_l$ and $U_{\nu}$ being the unitary matrices which diagonalize 
the $3\times 3$ charged lepton and neutrino mass matrices, 
$M_l$ and $M_{\nu}$, 
\begin{eqnarray}
U_l^{\dagger}M_l^{\dagger}M_lU_l &=& 
{\rm diag}(m^2_e,m^2_{\mu},m^2_{\tau}),  \\
U_{\nu}^{\dagger}M_{\nu}^{\dagger}M_{\nu}U_{\nu} &=& 
{\rm diag}(m^2_{\nu_e},m^2_{\nu_{\mu}},m^2_{\nu_{\tau}}), 
\label{mixing}
\end{eqnarray}
respectively.  
Here the left-handed neutrino mass matrix is expressed in terms of 
right-handed Majorana neutrino mass matrix, $M_R$, and Dirac neutrino mass 
matrix, $M_{\nu_D}$,
\begin{eqnarray}
     M_{\nu}=M_{\nu_D}^T M_R^{-1} M_{\nu_D}. 
\label{seesaw}
\end{eqnarray}
If we restrict ourselves to the case in which such large mixings are naturally 
derived without fine tuning, the origin of each of 
the large mixing angles, $\theta_{\mu \tau}$ and  $\theta_{e \mu}$, 
must be due to either $M_{\nu}$ or $M_l$.

In this paper, we show that these two large mixing angles can be 
derived from symmetric four zero 
texture within the $SO(10)$ GUT framework. 
We assume the following~~textures~~for~~up-~~and~~down-type~~quark mass 
matrices at the GUT scale~\cite{Nishiura},
\footnote{Here we neglect the CP phases, since they have little effect on 
the final result. The simple expressions Eq. (\ref{Md}) 
and (\ref{Mu}) are derived by 
using $m_u<< m_c<<m_t, \, m_d<<m_s<<m_b$.}
\bea
M_D &=& \left(
\begin{array}{@{\,}ccc@{\,}}
0 & \sqrt{\frac{m_d m_s m_b}{m_b-m_d}} & 0 \\ 
\sqrt{\frac{m_d m_s m_b}{m_b-m_d}} & 
m_s & \sqrt{\frac{m_d m_b (m_b-m_s-m_d)}{m_b-m_d}} \\ 
0 & \sqrt{\frac{m_d m_b (m_b-m_s-m_d)}{m_b-m_d}} & m_b-m_d 
\end{array}
\right) \nonumber \\
&\simeq& m_b  \left(
\begin{array}{@{\,}ccc@{\,}}
0 & \frac{\sqrt{m_d m_s}}{m_b} & 0 \\ 
\frac{\sqrt{m_d m_s}}{m_b} &\frac{ m_s}{m_b} &  \sqrt{\frac{m_d}{ m_b}} \\ 
0 & \sqrt{\frac{m_d}{ m_b}} &1
\end{array}
\right) 
\label{Md}
\eea
\bea
M_U 
&\simeq& m_t \left(
\begin{array}{@{\,}ccc@{\,}}
0 & \frac{\sqrt{m_u m_c}}{m_t} & 0 \\ 
 \frac{\sqrt{m_u m_c}}{m_t} & \frac{m_c}{m_t} &  \sqrt{\frac{m_u}{ m_t}} \\ 
0 & \sqrt{\frac{m_u}{ m_t}} & 1 
\end{array}
\right)
\equiv m_t\left(
\begin{array}{@{\,}ccc@{\,}}
0 & a & 0 \\ 
a & b & c \\ 
0 & c & 1
\end{array}
\right).
\label{Mu}
\eea
As for $M_D$ it is well known that each elements of 
$M_U$ and $M_D$ is dominated by the contirbution  either 
from ${\bf 10}$ or ${\bf 126}$ Higgs fields, 
where the ratio of Yukawa couplings of charged lepton to down quark 
are $1$ or $-3$, respectively. 
More concretely the following option for $M_D$ (Georgi-Jarlskog type~\cite{GJ}) 
\begin{eqnarray}
M_D =
\left(
\begin{array}{@{\,}ccc@{\,}}
 0                 &{\bf 10}           & 0   \\
{\bf 10}           &{\bf 126}          &{\bf 10} \\
 0                 &{\bf 10}           & {\bf 10}
\end{array}\right), 
\eea
is known to reproduce very beautifully all the experimental data of 
$m_{\tau}, m_{\mu}, m_e$ as well as $m_b, m_s, m_d$~\cite{HRR,DHR}. 
On the other hand, $M_U$ is related to $M_{\nu_D}$, 
which is not directly connected to neutrino experiments 
no definite configuration has been found so far. 
Here we show the following option  
reproduces two large mixings at the same time, 
\bea
M_U =
\left(
\begin{array}{@{\,}ccc@{\,}}
 0                 &{\bf 126}           & 0   \\
{\bf 126}           &{\bf 10}            &{\bf 10} \\
 0                 &{\bf 10}            & {\bf 126}
\end{array}\right),
\label{126up}
\end{eqnarray}
which uniqely determines neutrino Dirac mass matrix as
\bea
M_{\nu_D} = m_t 
\left(
\begin{array}{@{\,}ccc@{\,}}
0 & -3a & 0 \\ 
-3a & b & c \\ 
0 & c & -3
\end{array}
\right).
\eea
For the right-handed Majorana mass matrix, to which only 
${\bf 126}$ Higgs field couples, we assume the following 
option consistently with $M_U$:
\bea
M_R
\equiv m_R 
\left(
\begin{array}{@{\,}ccc@{\,}}
0 & r  & 0 \\ 
r   & 0 & 0 \\ 
0   & 0 & 1
\end{array}
\right).
\eea
Then from  Eq.~(\ref{seesaw}), we get 
\bea
M_{\nu}
= \left(
\begin{array}{@{\,}ccc@{\,}}
 0  & \frac{a^2}{r} & 0   \\
\frac{a^2}{r} & \frac{-2a b}{3r} +\frac{ c^2}{9}
& -\frac{ c}{3} (\frac{a}{r}+1) \\
 0  &-\frac{ c}{3} (\frac{a}{r}+1) & 1
\end{array}
\right) \frac{9m_t^2}{m_R} . 
\label{nuabcr}
\eea
Since the order of the parameters in the above Eq.~(\ref{nuabcr}) are 
$a<<  b\leq c<<1$, we recognize that the first term of the 3-2 element,  
$ \frac{-3ac}{r}$,  must be of order 1 in order to get large 
mixing angle $\theta_{\mu \tau}$, so we here take   
\begin{equation}
r \simeq \frac{ac}{3} \simeq  
 \frac{\sqrt{m_u^2m_c}}{3\sqrt{m_t^3}}
\sim10^{-7}. 
\label{eq:mjratio}
\end{equation}
This is almost the same situation 
as discussed by Kugo, Yoshioka and one of 
the present authors~\cite{Bando:1997ns}. 
This tiny value of $r$ is very welcome~\cite{Bando:1997ns}; 
the right-handed Majorana mass of 
the third generation  must become of the order of GUT scale while those of 
the first and second generations are of order $10^{8}$ GeV. 
This is quite favorable for the
GUT scenario to reproduce the bottom-tau mass ratio. 

Up to here the situation is quite trivial in a sense;  
one arbitrary parameter $r$ has been chosen so as to reproduce 
the large mixing angle $\theta_{\mu\tau}$. 
Now the problem is whether it naturally reproduces 
another mixing angle 
$\theta_{e\mu}$. 
At this stage we have no arbitrary parameter to 
adjust the mass ratios or mixing angles. 
Under such condition, $M_{\nu}$ is approximately written as 
\begin{eqnarray}
M_{\nu}=
\left(
\begin{array}{@{\,}ccc@{\,}}
 0               & \frac{-3a}{c}  & 0   \\
 \frac{-3a}{c}   & \frac{2b}{c}   & -1 \\
 0  & -1 & 1
\end{array}
\right) \frac{9m_t^2}{m_R}
\equiv
\left(
\begin{array}{@{\,}ccc@{\,}}
 0      & \beta  & 0   \\
\beta   &\alpha   & -1 \\
 0  & -1 & 1
\end{array}
\right) \frac{9m_t^2}{m_R}, 
\label{apmnu}
\end{eqnarray}
with $\alpha=\frac{2b}{c}=\frac{2m_c}{\sqrt{m_tm_u}} $ and 
$\beta=\frac{-3a}{c}=-3\sqrt{\frac{m_c}{m_t}}$.  
Since $\beta<<\alpha \sim 1 $, we first diagonalize the $2 \times 2$ 
matrix of the 2-3 block of Eq. (\ref{apmnu}) by rotating the following angle 
\bea
\tan^22\theta_{\mu\tau}= \frac{4}{(1-\alpha)^2},
\eea
through which $M_{\nu}$ is deformed as follows 
\begin{eqnarray}
\left(
\begin{array}{@{\,}ccc@{\,}}
 0      & \beta  & 0   \\
\beta   &\alpha   & -1 \\
 0  & -1 & 1
\end{array}
\right) \quad \rightarrow \quad 
\left(
\begin{array}{@{\,}ccc@{\,}}
0 &\beta \cos\theta_{\mu\tau} &\beta \sin\theta_{\mu\tau} \\
\beta \cos\theta_{\mu\tau}  &\lambda_2  &0 \\
\beta \sin\theta_{\mu\tau}  & 0         &\lambda_3
\end{array}\right),
\label{mnu23}
\end{eqnarray}
with 
\begin{eqnarray}
\lambda_3 &=& \frac{\alpha+1+\sqrt{(\alpha-1)^2+4}}{2} \, \,
(\simeq \, \lambda_{\tau}), \\ 
\lambda_2 &=& \frac{\alpha+1-\sqrt{(\alpha-1)^2+4}}{2}.
\label{eigenvaluelam}
\end{eqnarray}
Then we further diagonalize the $2\times 2$ matrix of the 1-2 block 
of Eq. (\ref{mnu23}) by rotating $\theta_{e \mu}$:
\begin{eqnarray}
M_{\nu}
\rightarrow 
\left(
\begin{array}{@{\,}ccc@{\,}}
\lambda_e      & 0  &\beta \sin\theta_{\mu\tau}\cos\theta_{e \mu}   \\
0   &\lambda_{\mu}  & \beta \sin\theta_{\mu\tau}\sin\theta_{e \mu}   \\
\beta \sin\theta_{\mu\tau}\cos\theta_{e \mu}  
& \beta \sin\theta_{\mu\tau}\sin\theta_{e \mu}     & \lambda_{\tau}
\end{array}
\right),  
\end{eqnarray}
with the rotating angle 
\bea
\tan^2 2\theta_{e \mu}
= \Biggl( \frac{2\beta \cos\theta_{\mu\tau}}{\lambda_2} \Biggr)^2, 
\label{tanemu}
\eea
and mass eigenvalues
\begin{eqnarray}
\lambda_{\mu} &=& 
\frac{\lambda_2+\sqrt{\lambda_2^2+4\beta^2\cos^2\theta_{\mu\tau}}}{2}, \\
\lambda_e &=&
\frac{\lambda_2-\sqrt{\lambda_2^2+4\beta^2\cos^2\theta_{\mu\tau}}}{2}.
\label{lambdadasshu}
\end{eqnarray}
Finally we get the following approximate form for the rest 
small rotating angle, 
\begin{equation}
\sin \theta_{e3} = \frac{2\beta \sin\theta_{\mu\tau}\cos \theta_{e \mu}}
{\lambda_{\tau}}.   
\end{equation}
Leaving details in a separate paper~\cite{BO2}, we demonstrate how we can 
predict neutrino masses and mixings; 
all the neutrino information are determined 
in terms of $m_u,m_c,m_t$ as
\begin{eqnarray}
\tan^2 2\theta_{\mu\tau} &\simeq& 
\frac{1}{(1-\frac{2m_c}{\sqrt{m_um_t}})^2},  \\
\tan^2 2\theta_{e\mu} &\simeq&
\frac{9m_c}{m_t(1-\frac{2m_c}{\sqrt{m_um_t}})^2}, \\   
\sin \theta_{e3}  &\simeq& 
-3 \sqrt{\frac{m_c}{m_t}} \sin\theta_{\mu\tau}\cos \theta_{e \mu}, 
\end{eqnarray}
from which the following equations are derived
\bea
\tan^22\theta_{e \mu} \simeq \frac{9m_c}{m_t}\tan^22\theta_{\mu\tau},
\qquad 
\sin^2 \theta_{e3}  \simeq  
\frac{9m_c}{m_t}\sin^2 \theta_{\mu\tau}\cos^2\theta_{e \mu}.
\eea
This indecates that $\tan^22\theta_{e \mu}$ is smaller 
by a factor $\frac{9m_c}{m_t}$ than $\tan^22\theta_{\mu\tau}$. 
Intersting enough is that once we know the atmospheric and 
solar neutrino experiments, $U_{e3}$ is predicted without 
any ambiguity coming from the up-quark masses at GUT scale;  
\bea
\sin^2 \theta_{e3}  \simeq 
\frac{\tan^22\theta_{e \mu}}{\tan^22\theta_{\mu \tau}} \cdot
\sin^2 \theta_{\mu \tau}\cos^2\theta_{e \mu},
\eea
which is independent of the uncertainty especially coming from the value, 
$m_t$, at GUT scale. 

Next the neutrino masses are given by
\bea
m_{\nu_{\tau}} \simeq \lambda_{\tau} \cdot  \frac{m_t^2}{m_R},\quad 
m_{\nu_{\mu}} \simeq  \lambda_{\mu} \cdot  \frac{m_t^2}{m_R},\quad 
m_{\nu_e} \simeq \lambda_e \cdot \frac{m_t^2}{m_R}, 
\label{neutmass}
\eea
where the renormalization factor ($\sim \frac{1}{3}$) has been 
taken account 
to estimate the lepton masses at low energy scale. 
Since $\lambda_{\mu} << \lambda_{\tau} \sim O(1)$, 
this indeed yields $m_R\sim 10^{16}$ GeV, as many people 
require. 
On the other hand, $m_{\nu_{\mu}}$ is expected to become 
almost of the same order to $m_{\nu_e}$ as is seen 
from Eq.~(\ref{neutmass}). 

\begin{figure}[b]
\begin{center}
\includegraphics[width=8cm,clip]{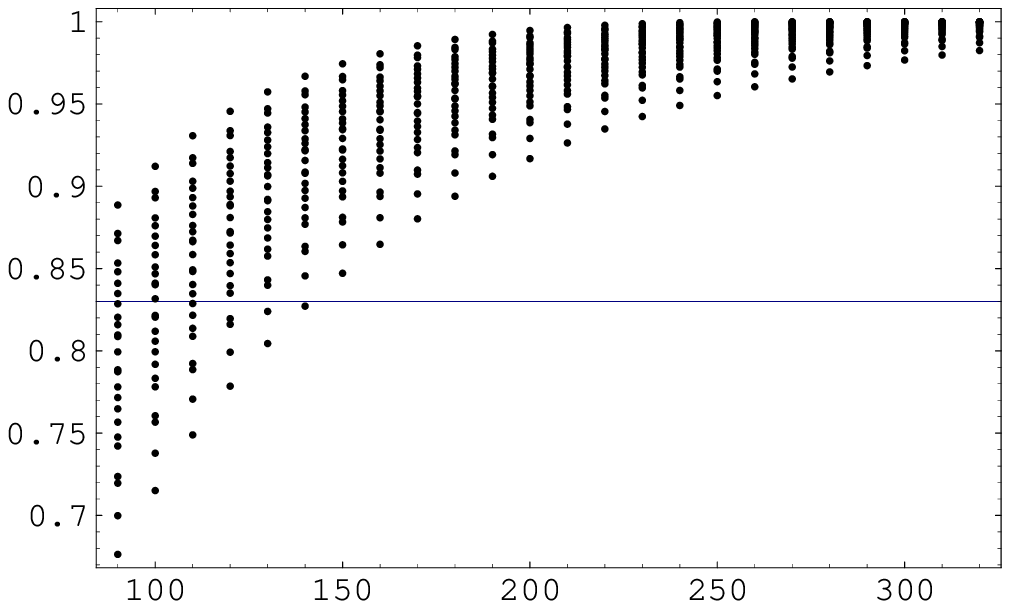}
\end{center}
\caption{Calculated values of $\sin^2 2\theta_{\mu \tau}$ 
versus $m_t$. The parameter region of  $m_t$ at GUT scale is within 
the  allowed uncertainty. The horizontal line 
indecates the experimental lower bound for atmospheric neutrinos.} 
\label{fig:top-sin22}
\end{figure}

\begin{figure}[t]
\begin{center}
\includegraphics[width=8cm,clip]{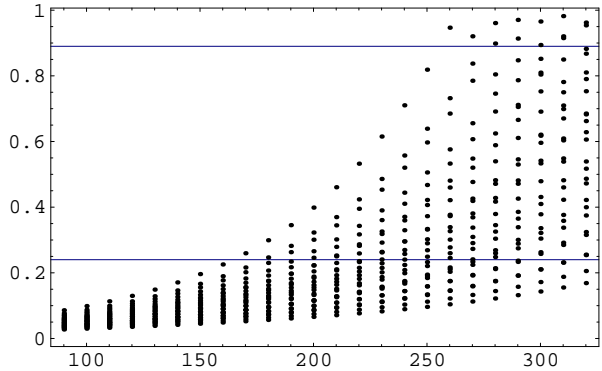}
\end{center}
\caption{$\tan^2 \theta_{e \mu}$ as function of the 
top quark mass, $m_t$. 
The parameter region of $m_t$ is taken 
within the uncertainty of $m_t$ at GUT scale. 
The horizontal lines indicalte the experimental allowed regions 
for solar neutrinos.}
\label{fig:top-tan2}
\end{figure}

\begin{figure}[t]
\begin{center}
\includegraphics[width=8cm,clip]{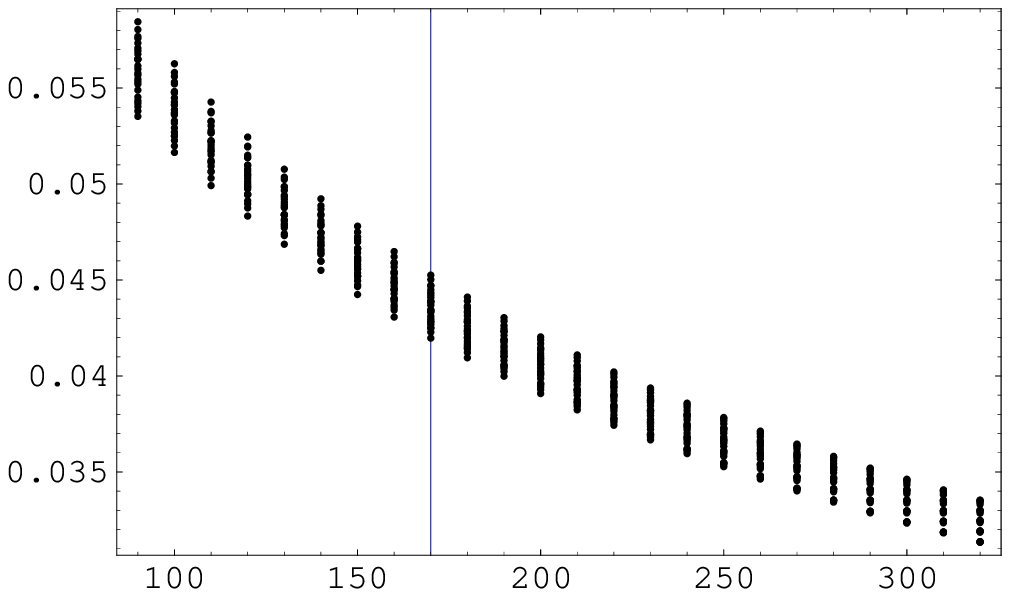}
\end{center}
\caption{Predicted values of $U_{e3}$ 
versus $m_t$ at GUT scale within uncertainty.  
The vertical line indicates the point $m_t = 170~\gev$.}
\label{fig:top-Ue3}
\end{figure}

Let us make some numerical calculation. We take the values  
of $m_t,m_c,m_u$ at GUT scale obtained by 
Fusaoka and Koinde~\cite{koide-fusaoka};
\bea
m_u &=& 1.04^{+0.19}_{-0.20}~\mev, \\
m_c &=& 302^{+25}_{-27}~\mev, \\
m_t &=& 129^{+196}_{-40}~\gev.
\eea
Among them $m_t$ is the most sensitive parameter.  
Fig.~\ref{fig:top-sin22} and 
Fig.~\ref{fig:top-tan2} 
show the dependence of the resultant values of 
$\theta_{\mu \tau}$ and $\theta_{e \mu}$ on $m_t$, respectively.  
From Fig.~\ref{fig:top-sin22} $m_t$ is found to be larger 
than $90~\gev$ and from Fig.~\ref{fig:top-tan2} the lower bound is 
$m_t=170~\gev$.  With this bound for $m_t~(170-320~\gev)$ we can 
predict the values of $U_{e3}$,
\bea
U_{e3} \sim 0.03-0.045.  
\eea
from Fig.~\ref{fig:top-Ue3}.
We hope this can be checked by experiment in near future
JHF-Kamioka long-base line~\cite{JHF}, the sensitivity of 
which is reported as $|U_{e3}|\simeq 0.04~\rm at~90~ \%~ C.L.$. 
If we further expect 
Hyper-Kamiokande ($|U_{e3}|<10^{-2}$)~\cite{Hyper-K}, 
we can completely check 
whether such symmetric texture model can survive or not. 
In conclusion we list a set of typical values of neutrino masses and 
mixings at $m_t \simeq 240~\gev$; 
\bea
\sin^2 2\theta_{\mu \tau} &\sim& 0.95-1, \\ 
\tan^2 \theta_{e \mu} &\sim& 0.23-0.6, \\
U_{e3}  &\sim& 0.037-0.038, \\
m_{\nu_{\tau}} &\sim& 0.06-0.07~\rm eV, \\
m_{\nu_{\mu}} &\sim& 0.003-0.006~\rm eV, \\
m_{\nu_e} &\sim& 0.0007-0.0015~\rm eV,
\eea
with $m_R= 2 \times 10^{15}~\gev$ and $r m_R=10^{8}~\gev$, which 
correspond to the Majorana mass for the third generation and 
those of the second and first generations, respectively.

Remark that, once the 
scale of right-handed Majorana mass matrix, $r$,  
is determined so as for the mixing angle of atmospheric neutrino 
to become maximal, the same vale $r$ well reproduces  
the ratio of 
the mass differences  
$\Delta m^2_{\mu \tau}$ to $\Delta m^2_{e \mu}$. 

We add two comments. 
First, one might suspect that we may always reproduce any desired 
neutrino mass matrix by adjusting the 
parameters appearing  in $M_R$, namely we can take 
$M_R = M_{\nu_D} M_{\nu}^{-1} M_{\nu_D}^T$. 
However this is not actually true if the mass matrix 
$M_U$ is of  such hierarchical structure 
as those coming from the famous anomalous $U(1)$. 
We shall show this in a separate paper~\cite{BO2}, 
where other options  of 
Eq.~(\ref{126up}) are fully investigated to confirm that 
only the option adopted in this paper can reproduce 
the two large mixing angles. 
Second, our scenario is quite differnt from those 
starting from the assumption 
that the large mixing angles observed in neutrino 
oscillation data comes from the charged lepton mass matrix; 
one might predict some relations of 
neutrino mixing angles to down quark information~\cite{Bando:2000at}, 
but we would  no more predict
the absolute vales of neutrino masses, which indeed 
needs the information of $M_{\nu}$. 
Our scenario, if it is indeed true, can 
predict without any ambiguity even for the order-one coefficients. 
The remarkable results are obtained really thanks to the power of GUT.
\section*{Acknowledgements}

This work started from the discussion 
at the research meeting held in Nov. 
2002 supported by the Grant-in Aid for Scientific Research
No. 09640375. 
We would like to thank to  A.~Sugamoto and T.~Kugo 
whose stimulating discussion encouraged us very much. 
Also we are stimulated by  the fruitful and instructive discussions 
during the Summer Institute 2002 held at 
Fuji-Yoshida. 
M.~B.\  is supported in part by
the Grant-in-Aid for Scientific Research 
Nos.~12640295 from Japan Society for the Promotion of Science, and 
Grants-in-Aid for Scientific Research on Priority Area A 
``Neutrinos" (Y.~Suzuki) Nos.~12047225, 
from the Ministry of Education, Science, Sports and Culture, Japan.
Also we are thankful to T.~Takeuchi for his kind comment 
on our poor English. 
%
%
%
%

\end{document}